\documentstyle[12pt,epsfig]{article}
\advance \topmargin by -\headheight
\advance \topmargin by -\headsep

\evensidemargin \oddsidemargin
\begin{document}
\topmargin -.6in
\def\br{\begin{eqnarray}}
\def\er{\end{eqnarray}}
\def\be{\begin{equation}}
\def\ee{\end{equation}}
\def\({\left(}
\def\){\right)}
\def\a{\alpha}
\def\b{\beta}
\def\d{\delta}
\def\D{\Delta}
\def\g{\gamma}
\def\G{\Gamma}
\def\h{ {1\over 2}  }
\def\hp{ {+{1\over 2}}  }
\def\hm{ {-{1\over 2}}  }
\def\k{\kappa}
\def\l{\lambda}
\def\L{\Lambda}
\def\m{\mu}
\def\n{\nu}
\def\o{\over}
\def\O{\Omega}
\def\p{\phi}
\def\rh{\rho}
\def\s{\sigma}
\def\t{\tau}
\def\th{\theta}
\def\ii {\'\i  }

\begin{center}   {\large {\bf  Supersymmetric variational energies of 3d confined
potentials}}\footnotemark 
\end{center} 
\normalsize 
\vskip 1cm 
\begin{center} {\it  Elso Drigo Filho} 
\footnotemark \footnotetext{elso@df.ibilce.unesp.br}  \footnotemark \footnotetext{Work
partially supported by CNPq} \\ 
Instituto de Bioci\^encias, Letras e  Ci\^encias Exatas,
IBILCE-UNESP\\ 
Rua Cristov\~ao Colombo, 2265 -  15054-000 S\~ao Jos\'e do Rio Preto - SP\\ 
\vskip 1cm 
{\it Regina Maria  Ricotta } \footnotemark \footnotetext{regina@fatecsp.br}\\ Faculdade de
Tecnologia de S\~ao Paulo, FATEC/SP-CEETPS-UNESP \\ 
Pra\c ca  Fernando Prestes, 30 -  01124-060 S\~ao Paulo-SP\\ Brazil\\ 
\vskip 1cm 
 
{\bf  Abstract}\\  
\end{center}
Within the approach of Supersymmetric Quantum Mechanics associated with the variational  method a
recipe to construct the superpotential of three dimensional confined potentials in general is
proposed. To illustrate the construction, the energies of the Harmonic Oscillator and the
Hulth\'en potential, both confined in three dimensions are evaluated. Comparison with the
corresponding results of other approximative and exact numerical results is presented. 
 
\newpage
\noindent {\bf 1. Introduction}\\ 

Supersymmetric Quantum Mechanics, (SQM),  was conceived in   1981  as the simplest
field theory  that would provide understanding of the supersymmetry breaking mechanism of higher
dimensional field theories, \cite{Witten}. Since then this formalism has become a new field of
research and a great deal of work, both analytical and numerical, has been done  to get better
knowledge  of  the exactly solvable, the partially solvable, the isospectral, the periodic and
the non-exactly solvable  potential problems, (see \cite{Cooper} for the latest review). In
particular, few of these latter have been investigated through a novel methodology based on the
association of the variational method with SQM formalism.  The conjunction of this well known
quantum mechanism with SQM has given a simple framework to investigate spectral problems of
Hamiltonian systems associated to the three dimensional Hulth\'en, Morse and Screened Coulomb
potentials,\cite{Drigo1}-\cite{Drigo3}. The main basis of the method consists in setting an {\it Ansatz} for the superpotential. Based on the superalgebra intrinsic of SQM we are able to evaluate  the wavefunction which naturally contains free parameters introduced by the {\it Ansatz}.  Having in hands our ideal wavefunction, the calculations to be done are the usual variational calculations in which the parameters are varied until the energy expectation value reaches its minimum.     

 More recently the application of this methodology to
exploit the energy spectra of a confined quantum mechanical system gave encouraging results: the
confined 3-dimensional hydrogen atom showed results compatible with those obtained from other
approximative methods  and numerical calculation, \cite{Drigo4}. 

Motivated by these results and considering the widespread interest in quantum confined systems, particularly in the studies of  semiconductor heterostructures (nanostructures like quantum dots, quantum wires, quantum wells), \cite{Kostov}-\cite{Yildirim}, as well as in field theory, \cite{Castro}, in this work we propose a systematic construction of the superpotential for 3d confined systems, that yields, through SQM, to the trial wavefunction pursued.

To illustrate the construction, the technique is applied here to get the $1s$ and $2p$ energy states for the confined Harmonic Oscillator and for the $1s$, $2p$ and $3d$ energy states of the confined Hulth\'en, both in three dimensions. They are very good when compared to recent results obtained from other approximative methods as well as exact numerical  results, \cite{Aguilera}-\cite{Sinha2}. 

In what follows, in order to be self consistent and to set up the notation, we present  a brief description of SQM and its association with the variational method. At this point we introduce our recipe for the superpotential in the general case.  Then we show the results of the application of this formalism to the confined Harmonic Oscillator and Hulth\'en potential. Comparison with the corresponding results of  other approximative and exact numerical results is presented in the tables,  followed by a discussion and our conclusions.\\  

\noindent {\bf 2. Supersymmetric Quantum Mechanics and the Variational Method}\\  

The starting point is the factorization of a Hamiltonian $H$ associated to a given
potential $V(r)$, in terms of bosonic operators $A^{\pm}$ and the lowest energy eigenvalue $
E_0$. In $\hbar = c = 1$ units, this is written as 
\be \label{H} H =  - \h{d^2 \o d r^2} + V(r) =  A^+A^-  + E_0 .
\ee 

We recall that for the cases we are considering the potential is symmetric so that equation 
(\ref{H}) is the radial equation and  $V(r)$ includes the barrier potential term. The bosonic
operators are  defined in terms of the so called superpotential
$W(r)$, 
\be A_1^{\pm} =  {1\o \sqrt 2}\left(\mp {d \o dr} + W(r) \right) .
\ee 

Thus, substituting the bosonic operators into equation (\ref{H}) we conclude that,  as a
consequence of the factorization of the Hamiltonian $H$, the Riccati equation must be satisfied, 
\be
\label{Riccati} W^2 - W'= 2\left( V(r) - E_0\right), 
\ee 
where $W' = {dW \o dr}$.  Also through
the superalgebra, the eigenfunction for the lowest state is related to the superpotential $W$ by 
\be \label{eigenfunction} \Psi_0(r) = N exp( -\int_0^r W(\bar r) d\bar r).  
\ee 

The above scheme works perfectly well if the potential is exactly solvable. However,  if the potential is
non-exactly solvable, the Hamiltonian is not exactly factorizable,  i.e., there is no 
superpotential that satisfies equation (\ref{Riccati}). On the other hand, the Hamiltonian can be
factorized in terms of an approximated  superpotential giving rise to an effective potential. This
is achieved by making an {\it Ansatz} for the superpotential, which naturally introduces free
parameters in the problem, denoted by the set $\{\mu\}$. Thus, through the superalgebra, the
wavefunction, (equation (\ref{eigenfunction})),  is evaluated. It is the trial wavefunction of
the usual variational method, denoted by $ \Psi_{\mu}$ and depending on 
$\{\mu\}$. Therefore, having in hands the main ingredient of the variational method, namely the 
trial wavefunction, the energy expectation value given by   
\be \label{energymu} E_{\mu} =
{\int{\Psi_{\mu}^* H \Psi_{\mu} dr}\over {\int{\mid \Psi_{\mu} \mid^2 dr}}} 
\ee  
can be evaluated and the parameters  can be  varied until it reaches its minimum value, $E(\bar\mu)$,
which is an upper limit of the sought energy level.  Indeed,  the  {\it Ansatz} in the
superpotential drives us to an effective potential $V_{eff}$ that has a similar form as the
original potential, i.e., 
\be 
\label{effective} V_{eff}(r) = \h\left(\bar W^2 - \bar W'\right)+ E(\bar\mu) 
\ee
where $ \bar W = W(\bar\mu)$ is the superpotential for $\{\mu = \bar\mu\}$, the set of
parameters that  minimises (\ref{energymu}).  The above methodology was applied in the search of
approximate energy levels of 3-dimensional systems,  like the Hulth\'en, Morse and Screened
Coulomb potentials, \cite{Drigo1}-\cite{Drigo3}. 

Concerning the confinement, it is remarkable that for the confined
cases we have treated so far, the confined Coulomb, \cite{Drigo4},  and the present cases of the
confined  Oscillator and Hulth\'en potentials,  the {\it Ansatz} set in the
superpotential is restricted to an extra confining term added to the superpotential used in the
non-confined case, i.e., 
\be \label{Wconf} W_{confined} = W_{non-confined} + {constant \o R -
r} 
\ee 
where $R$ is the radius of confinement. The advantage of this choice is that it allows us a
previous comparative analysis between the original potential (non-confined) and the effective
potential, containing the infinite (confining) barrier. This is corroborated by the fact that in the $R\rightarrow \infty$ limit the non-confined superpotential is recovered.   Moreover, the trial wavefunction will naturally vanish at the barrier $R = r$ and not as an input, as it is usual in the  variational calculation. \\

\noindent   {\bf  3. The Confined Harmonic Oscillator}\\

The radial Hamiltonian equation for the Harmonic Oscillator potential, written in atomic units is
given by

\be \label{Hamiltonian1} H = - \h {d^2 \o dr^2}  + {l(l+1) \o 2r^2} + {r^2 \o 2}.
\ee  

As the Harmonic Oscillator potential is symmetric, the confinement is introduced by an infinite potential
barrier at radius $r = R$. Thus we make the following  {\it Ansatz} for the superpotential 
\be
\label{W1}  W(r) =  - {\mu_1 \o r} + {\mu_2 \o R - r} + \mu_3 r 
\ee
which depends of $R$, the radius of confinement, and of three variational parameters, $\mu_1$,
$\mu_2$ and $\mu_3$. As stated in expression (\ref{Wconf}), the first and the last terms are
already known from the non-confined case, \cite{Cooper}. The second term deals with the
confinement. This is explicitly shown through the effective potential. From equation 
(\ref{effective}), it has the following form 
\br 
\label{VeffHO}
V_{eff}(r) & = & {\bar\mu_3\o 2} r^2 +
{\bar\mu_1(\bar\mu_1 - 1) \o 2 r^2} - \bar\mu_1 
\bar\mu_3 - {\bar\mu_3 \o 2} +  \nonumber \\ & &{\bar\mu_2(\bar\mu_2 - 1) \over 2 (R - r)^2} -  
{\bar\mu_1 \bar\mu_2 \o r(R - r)} + {\bar\mu_2 \bar\mu_3 r \o (R - r)} +  E(\bar\mu)    
\er  which is clearly infinite at $r = R$, as expected for a confining system. Notice that the
effective potential is evaluated for  the values of the set of parameters $\{\bar\mu\}$ that
minimise the energy.  Our trial wavefunction for the variational method is obtained from the
superalgebra through equation  (\ref{eigenfunction}), using the superpotential given by the {\it
Ansatz} made in equation (\ref{W1}). It is given by
\be \label{Psi-HO}
\Psi_{\mu}(r) = \Psi(\mu_1, \mu_2, \mu_3, r) \propto  r^{\mu_1} \; (R - r)^{\mu_2} \;e^{-\mu_3 
r^2/2}.
\ee 

It depends of  three free parameters, $\mu_1$,  $\mu_2$  and  $\mu_3$ and vanishes at $r =
R$.  We use this trial wavefunction to calculate the energy expection value, given by  the
equation (\ref{energymu}), evaluated with the Hamiltonian (\ref{Hamiltonian1}). Its  minimisation
with respect to the three parameters gives $E(\bar\mu)$, which, from now on, we call
$E_{vsqm}$.  The results are given in the tables below for different values of the confining
radius $R$ and states $1s$ and $2p$, corresponding to the values of $l = 1$ and $2$ of the 
angular momentum, respectively. Comparison is made with exact numerical, $E_{exact}$, 
perturbative,
$E_{pert}$, \cite{Aguilera}, other variational, $E_{var}$, \cite{Marin}, WKB, $E_{wkb}$,
\cite{Sinha1} and modified WKB results, $E_{centri}, $\cite{Sinha2}, 
through the percentage errors, \be \delta_{vsqm} = {|E_{exact}-E_{vsqm}|\over
E_{exact}}\% \ee and \be \delta_{pert} = {|E_{exact}-E_{pert}|\over
E_{exact}}\%. \ee 
The same percetage errors are applied to the other results. We must note that the
perturbative results lost their accuracy for large values of the radius of confinement, for
which  they are outside the convergence region. For the results of Table 1, $R < 1.72$ and of
the Table 2, $R < 1.84$. 

It should be stressed that in the limit of no-confinement, i. e. $R\rightarrow \infty$, the variational SQM results also agree with the exact 
non-confined problem, corresponding to the removal of the infinite barrier. In this
case the energy is exact and is given by 
\be E = 2n + L + 3/2\;.
\ee
Thus for $n = 0$ and $l = 0$, state $1s$, the exact result is $E_{R \rightarrow\infty} = 1.5000$,
(Table 1), and for
$n = 0$ and $l = 1$, state $2p$,  the exact result is $E_{R \rightarrow\infty} = 2.5000$, (Table
2). 
\vskip .3cm

\noindent {\bf Table 1.}  Energy eigenvalues (in Rydbergs) and percentage errors for different
values of $R$ for the $1s$ state, ($l=0$). \\
\vskip .3cm
\noindent \begin{tabular}{|r|c|c|c|c|c|c|c|c|c|} \hline
\multicolumn{1}{|c|} {R} & 
\multicolumn{1}{|c|} {$E_{exact}$} &
\multicolumn{1}{|c|} {$E_{vsqm}$} & 
\multicolumn{1}{|c|} {$\delta_{vsqm}$} &
\multicolumn{1}{|c|} {$E_{pert}$} &
\multicolumn{1}{|c|} {$\delta_{pert}$} & 
\multicolumn{1}{|c|} {$E_{wkb}$} &
\multicolumn{1}{|c|} {$\delta_{wkb}$} & 
\multicolumn{1}{|c|} {$E_{var}$} &
\multicolumn{1}{|c|} {$\delta_{var}$} \\ \hline 
1.0 & 5.0755  & 5.0865 & 0.22 & -& - &5.0627 & 0.25 & 5.1313 & 1.10\\ \hline 
1.5 & 2.5050  & 2.5104 & 0.22& 2.5046 & 0.02& 2.5082& 0.13& 2.5265& 0.86\\ \hline 
2.0 & 1.7648  & 1.7664 & 0.09&1.7588 & 0.34&1.9882 & 12.7&1.7739 &0.52\\ \hline 
2.5 & 1.5514  &1.5529 &0.10 & - & -& 1.5564 & 0.32&1.5567 &0.34\\ \hline 
3.0 & 1.5061  & 1.5069 & 0.05&1.1532 & 23.4&1.5061  &0.00& 1.5105  &0.29\\ \hline
4.0 & 1.5000  & 1.5002 & 0.01&- 4.341728 & -& 1.5000 &0.00 &1.5033& 0.22\\ \hline 
5.0 & 1.5000  & 1.5000 & 0.00&-48.076319 && 1.5000 & 0.00&1.5025& 0.17\\ \hline 
10.0& 1.5000  & 1.5000 & 0.00&-& -&1.5000 & 0.00& -&- \\ \hline
\end{tabular}\\
\newpage
\noindent {\bf Table 2.}  Energy eigenvalues (in Rydbergs) and percentage errors for different values of $R$
for the $2p$ state, ($l=1$). \\
\vskip .3cm

\noindent \begin{tabular}{|r|c|c|c|c|c|c|c|c|c|c|c|} \hline
\multicolumn{1}{|c|} {R} & 
\multicolumn{1}{|c|} {$E_{exact}$} &
\multicolumn{1}{|c|} {$E_{vsqm}$} & 
\multicolumn{1}{|c|} {$\delta_{vsqm}$} &
\multicolumn{1}{|c|} {$E_{pert}$} &
\multicolumn{1}{|c|} {$\delta_{pert}$} & 
\multicolumn{1}{|c|} {$E_{wkb}$} &
\multicolumn{1}{|c|} {$\delta_{wkb}$} & 
\multicolumn{1}{|c|} {$E_{var}$} &
\multicolumn{1}{|c|} {$\delta_{var}$} &
\multicolumn{1}{|c|} {$E_{centri}$} &
\multicolumn{1}{|c|} {$\delta_{centri}$} \\ \hline 

0.3 &112.188 & 112.231& 0.04& 112.188& 0.00& -& & - & -& & -\\ \hline 
1.0 &10.2822 & 10.2847 & 0.02&- & -& 10.2643 & 0.17&10.3188 & 0.36& 10.2876& 0.05 \\ \hline 
1.5 & 4.9036 & 4.9046 &0.02&4.9034 & 0.00&4.9084& 0.01& 4.9169 &0.27& 4.9068& 0.07\\ \hline 
2.0 & 3.2469 & 3.2471 &0.01 &3.2434 &0.11 &3.2490& 0.07& 3.2514 &0.14& 3.3081& 1.88 \\ \hline 
2.5 &2.6881 & 2.6891 &0.04&- &-& 2.7079 &0.74& 2.6901&0.07 & 2.6835 &0.17\\ \hline 
3.0& 2.5313&2.5322 &0.04 &2.3104 &0.73 &2.5310 & 0.01&2.5337 &0.10 &2.5313 &0.00 \\ \hline 
4.0 &2.5001 & 2.5003 & 0.01&-1.5656& 163&2.5001 &0.00&2.5015 & 0.06&2.5001 & 0.00\\ \hline 
5.0& 2.5000& 2.5000 & 0.00&-34.0358& -&2.5000 &0.00 &2.5012 &0.05 &2.5000 & 0.00\\ \hline 
10.0& 2.5000 & 2.5000&0.00 &-&- &2.5000 &0.00& -&- & -&- \\ \hline
\end{tabular}\\
\vskip .3cm
\noindent   {\bf  4. The Confined Hulth\'en Potential }\\

The radial Hamiltonian equation for the Hulth\'en potential, written in atomic units is given by

\be \label{Hamiltonian2} H = - \h {d^2 \o dr^2}  + {l(l+1) \o 2r^2} - {\d e^{-\d r}\o 1-e^{-\d r}} 
\ee 

Once again, as the potential is symmetric, the confinement is introduced by the infinite 
potential barrier at radius $r = R$. The associated {\it Ansatz} for the superpotential follows
the equation (\ref{Wconf}) and the expression for the superpotential given for the unconfined case, as in
\cite{Drigo1}, 
 
\be \label{W2} W(r) = B_1 {e^{-\mu_1 r}\o (1 - e^{-\mu_1 r})} + C_1 +
{\mu_2 \o R-r} 
\ee 
where   
\be  \label{B}  B_1 = -\mu_1 (l+1) \;,\;\;\;\;C_1 = -\h
\mu_1 + {1\o l+1}.
\ee    

This superpotential gives rise through the superalgebra, (\ref{eigenfunction}), to the  following  trial wavefunction
\be
\label{Psi-Hulthen}
\Psi_{\mu} = \Psi(\mu_1, \mu_2 ,r) \propto {(1 - e^{-\mu_1 r})}^{-{B_1 \o {\mu_1}}}\;  \; e^{-
C_1 r}(R - r)^{\mu_2}.
\ee  

It depends of  two free parameters, $\mu_1$  and  $\mu_2$ and vanishes at $r = R$, as
expected,  since the effective potential, evaluated by using equations (\ref{effective}),
(\ref{W2}) and (\ref{B}), is infinite at this point. Its general form is  
\br 
\label{VeffHu}
V_{eff}(r) &=& -
{\bar  \mu_1 e^{-\bar  \mu_1 r}\o 1-e^{-\bar  \mu_1 r}}  + {l(l+1)\o 2}{{\bar  \mu_1}^2 e^{-2\bar  \mu_1 r}\o (1-e^{-\bar  \mu_1
r})^2} + {\bar  \mu_2 (\bar  \mu_2 - 1) \o 2(R-r)^2}  - {(l+1) \bar  \mu_1 \bar  \mu_2 e^{-\bar  \mu_1 r} \o (R-r)(1-e^{-\bar  \mu_1r})} -  \nonumber \\ && 
 {\bar  \mu_1 \bar  \mu_2 \o 2(R-r)}  
 + {\bar  \mu_2 \o (l+1)(R-r)} +  \h(-{ \bar  \mu_1\o 2} + {1 \o l+1})^2 + E(\bar \mu).  
\er 
As in the previous case,   $\{\bar\mu\}$ is the set of parameters that minimise the
energy, given by $E(\bar\mu)$,  which we also call,  from now on,
$E_{vsqm}$.  The results are in the tables below, for different values of the parameter
$\d$ and the confining  radius $R$, for the states $1s$, $2p$ and $3d$, corresponding to the
values of $l = 1, 2$ and $3$ of the angular momentum, respectively. Comparison is made with exact
numerical, $E_{exact}$, from the 1/N approximation, $E_{1/N}$, \cite{Sinha3} and modified WKB
results, $E_{centri}$, \cite{Sinha2}. 

For $R\rightarrow \infty$ exact values of the energy are
only for the $l = 0$ case, and are given by 
\be E = -\h ({1\over n} - {n \delta \over 2})^2
\ee
Thus for $\delta = 0.1$ and $n=1$  the exact result is $E_{R \rightarrow\infty} =
-0.45125$, (Table 3).
\vskip .3cm

\noindent {\bf Table 3.}  Energy eigenvalues (in Rydbergs) for different values of
$R$,  $n=1$ and $l=0,1,2$, (states $1s$, $2p$ and $3d$) and 
$\d = 0.1$.\\

\noindent \begin{tabular}{|l|c|c|c|c|c|c|c|c|} \hline
\multicolumn{1}{|c|} {R} & \multicolumn{1}{|c|} {l} & 
\multicolumn{1}{|c|} {$E_{exact}$} &
\multicolumn{1}{|c|} {$E_{vsqm}$} & 
\multicolumn{1}{|c|} {$\delta_{vsqm}$} &
\multicolumn{1}{|c|}{$E_{1/N}$} &
\multicolumn{1}{|c|} {$\delta_{1/N}$} & 
\multicolumn{1}{|c|} {$E_{centri}$} &
\multicolumn{1}{|c|} {$\delta_{centri}$}\\ \hline 
6.0 & 0 & -0.45053 &-0.44945 & 0.24& -0.45109 &0.12 & &\\ \hline
    & 1 & -0.00865 &-0.00808 & 6.59&-0.00294 &66.0 &-0.00782&9.60 \\ \hline 
7.0 & 0 & -0.45111 & -0.45043 & 0.15&-0.45181 &0.16& &\\ \hline
    & 1 & -0.04069 &-0.04037 & 0.79&  -0.03324 &18.3 &  -0.03976 &2.29\\ \hline ¨
8.0 & 0 & -0.45122 &-0.45076& 0.10  &-0.45193 & 0.16&& \\ \hline
    & 1 & -0.05783 & -0.05762 & 0.36 & -0.05293 &8.47&-0.05510 &4.72\\ \hline 
9.0 & 0 & -0.45125 & -0.45090 & 0.08 &-0.45188 &0.14& & \\ \hline
    & 1 & -0.06728 & -0.06712 & 0.24& -0.06389& 5.04& -0.06612&1.72\\ \hline 
10.0 & 0 & -0.45125 & -0.45098 &0.06&-0.45179 &0.12& &\\ \hline
    & 1 & -0.07257 & -0.07243 &0.19& -0.07008  &3.43& -0.07196&0.84\\ \hline 
25.0 & 0 & -0.45125 & -0.45125 &0.00&-0.45131 &0.01& &\\ \hline
    & 1  & -0.07918 & -0.07915 & 0.04&-0.07920 &0.03& -0.07921&0.04\\ \hline
    & 2  & -0.01390 & -0.01380 &0.72 &-0.01332 &4.17 &-0.01381&0.65\\ \hline 
50.0 & 0 & -0.45125 & -0.45125 & 0.00&-0.45126 &0.00&&\\ \hline
    & 1 & -0.07918  & -0.07918 & 0.00&-0.07920 &0.03& -0.07920 &0.03\\ \hline
    & 2  & -0.01448 & -0.01448 & 0.00&-0.01450 &0.14& -0.01450& 0.14\\ \hline
 \end{tabular}\\

\vskip 1cm 
\noindent {\bf Table 4.}  Energy eigenvalues (in Rydbergs) and percentage errors for different
values of $R$ for the $2p$ state, ($l=1$) and 
$\d = 0.2$.\\

\vskip .3cm
\noindent \begin{tabular}{|l|c|c|c|c|c|c|c|} \hline
\multicolumn{1}{|c} {R} & 
\multicolumn{1}{|c|} {$E_{exact}$ } &
\multicolumn{1}{|c|} {$E_{vsqm}$} & 
\multicolumn{1}{|c|} {$\delta_{vsqm}$} &
\multicolumn{1}{|c|}{$E_{1/N}$} &
\multicolumn{1}{|c|} {$\delta_{1/N}$} & 
\multicolumn{1}{|c|} {$E_{centri}$ } &
\multicolumn{1}{|c|} {$\delta_{centri}$}\\ \hline 
8 & -0.01731 & -0.01708 & 1.33& -0.01242&28.3& -0.01607&7.16\\
\hline 9 &-0.02749 & -0.02731  &0.65& -0.02428&11.7 &-0.02612&4.98\\ \hline 
10 & -0.03339 & -0.03323 &0.48& -0.03118&6.62&-0.03389 &1.50\\ \hline 
25 & -0.04188 & -0.04178  &0.24&-0.04199&0.26&  -0.04192 &0.10\\ \hline 
50 &-0.04189&-0.04189 &0.00& -0.04196&0.17& -0.04191  &0.05\\ \hline
 \end{tabular}\\
\newpage
\vskip .5cm
{\bf 5. Discussion and conclusions}\\ 

The study of the two 3-dimensional confined quantum systems considered here was made through the mechanism based on the association of the variational method with SQM, using the recipe given to construct the superpotential, equation (\ref{Wconf}). For the two cases considered, namely, the Harmonic Oscillator and the Hulth\'en potential, the {\it Ansatz}, as given by equations (\ref{W1}) and (\ref{W2}), respectively, naturally introduced free parameters in the superpotential. Thus, using the superalgebra the wave functions were
evaluated, showing, of course, explicit dependence on these parameters, equations (\ref{Psi-HO})
and (\ref{Psi-Hulthen}), respectively. They were our optimal wave functions, used to calculate the energy expectation values, equation (\ref{energymu}), as in the usual variational method. The parameters were varied until this expectation value reached its minimum, giving rise to the
searched energy states.

The peculiar feature of the approach is that, when doing the {\it Ansatz} 
to the superpotential following the recipe of equation (\ref{Wconf}), we were able to evaluate the effective potential, equation (\ref{effective}), and also equations (\ref{VeffHO}) and (\ref{VeffHu}), which resemble the real, confining potential under consideration.  In both cases, the effective potential is surely infinite at the border, $r=R$, and as a result the wave function, evaluated through the superalgebra, equation (\ref{eigenfunction}), naturally vanishes at the border because it finds a  potential barrier that increases until becoming impenetrable at $r = R$. Furthermore, for increasing values of $R$ the border effects vanish and our effective potential becomes the original non-confined potential. The results for the energies improved, when compared with the corresponding results of other approximative and exact numerical results, converging to the results of the  non-confined system. On the other hand, for small values of $R$, stronger are the border effects, but even so very small deviation from the exact numerical results were found. 

We conclude that our main result is the recipe given for the superpotential when confinement takes place, equation (\ref{Wconf}).  The inclusion of the confining term  ${1 \over R - r}$ in the superpotential was crucial to describe border effects. This prescription showed to be appropriate to get good results for the cases considered so far and can be a first trial to exploit other confined systems through this method. It is remarkable that the approach was applied to different systems. Contrary to the Coulomb potential, \cite{Drigo4},  and the Harmonic Oscillator, the Hulth\'en potential is not exactly solvable in three dimensions. This makes our recipe good enough to work in the general case and  reinforces the fact that the association of the superalgebra of SQM with the variational method gives a simple and useful framework to investigate  confined systems in general.

\end{document}